\documentclass[12pt,a4paper,final]{iopart}
\usepackage{iopams}

\expandafter\let\csname equation*\endcsname\relax
\expandafter\let\csname endequation*\endcsname\relax

\usepackage{amsmath}
\usepackage{amsfonts}
\usepackage{amsmath}
\usepackage{amssymb}
\usepackage{amsthm}
\usepackage[utf8]{inputenc}
\usepackage{graphicx}
\usepackage{bm}
\usepackage{bbm}
\usepackage{cite}

\usepackage[breaklinks=true,colorlinks=true,linkcolor=blue,urlcolor=blue,citecolor=blue]{hyperref}

\newcommand{\defeq}{\mathrel{\mathop:}=}

\renewcommand{\H}{\mathcal{H}}
\newcommand{\G}{\bm{\Gamma}}
\newcommand{\Ga}{\bm{\Gamma}_A}
\newcommand{\Gb}{\bm{\Gamma}_B}

\newcommand{\IGb}{\int_{\mathcal{V}_B} d\Gb}
\newcommand{\iu}{\mathbbm{i}}

\newtheorem{theorem}{Theorem}
\newtheorem{lemma}{Lemma}

\begin{document}

\title{Closure of superstatistics}

\author[cor1]{Sergio Davis$^{1,2}$}
\address{$^1$Research Center in the Intersection in Plasma Physics, Matter and Complexity (P$^2$mc), Comisión Chilena de Energía Nuclear, Casilla 188-D, Santiago, Chile}
\address{$^2$Departamento de Física y Astronomía, Facultad de Ciencias Exactas, Universidad Andres Bello. Sazi\'e 2212, piso 7, 8370136, Santiago, Chile.}
\ead{sergio.davis@cchen.cl}

\begin{abstract}
Plasmas and other systems with long-range interactions are commonly found in non-equilibrium steady states that are outside traditional Boltzmann-Gibbs statistics, but can be described using 
generalized statistical mechanics frameworks such as superstatistics, where steady states are treated as superpositions of canonical ensembles under a temperature distribution. In this work we 
solve the problem of inferring the possible steady states of a composite system $AB$ where subsystem $A$ is described by superstatistics and $E_{AB} = E_A + E_B$. Our result establishes a closure 
property of superstatistics, namely that $A$ is described by superstatistics if and only if $AB$ and $B$ are also superstatistical with the same temperature distribution. Some consequences of this 
result are discussed, such as the impossibility of local thermal equilibrium (LTE) for additive subsystems in non-canonical steady states.
\end{abstract}

\section{Introduction}

Complex systems with long-range interactions, such as plasmas and self-gravitating systems, are known to present non-canonical statistical distributions, usually power laws~\cite{Du2007, Oka2018, 
Ourabah2020c, Yoon2020}. Among the frameworks that are used to explain these distributions, non-extensive statistical mechanics (also known as Tsallis statistics)~\cite{Tsallis2009c} and 
superstatistics~\cite{Beck2003, Beck2004} are the most widely used. Superstatistics seems one of the most promising, mostly because it is firmly grounded in probability theory and does not require 
additional assumptions. In particular, superstatistics obtains power-law distributions without replacing the Boltzmann-Gibbs entropy functional. Superstatistics is closely linked to the velocity 
distributions of particles in collisionless plasmas~\cite{Ourabah2015, Davis2019b, Ourabah2020b, Sanchez2021, Ourabah2024}, providing a plausible explanation for the origin of kappa 
distributions~\cite{Livadiotis2017, Lazar2021} in this and possible other kind of systems with long-range interactions.

An interesting challenge in non-equilibrium plasmas is to infer statistical properties of the system at larger scales from single-particle velocity distributions. As it has been previously 
shown~\cite{Davis2019b} that collisionless plasmas must have single-particle velocity distributions consistent with superstatistics, understanding the constraints on phase space distributions 
that allow for superstatistical velocity distributions is an important issue in advancing the statistical-mechanical theory of non-equilibrium plasmas.

In this work we prove a central result on the superstatistical behavior of composite systems, namely that for a composite system $AB$ with degrees of freedom $\bm{\Gamma} \defeq (\Ga; \Gb)$ divided into 
two subsystems $A$ and $B$ such that its Hamiltonian is \[\H_{AB}(\bm \Gamma) = \H_A(\Ga) + \H_B(\Gb),\] superstatistics \emph{propagates upwards}. That is, if the subsystem $A$ is superstatistical, 
then the composite system $AB$ must also be superstatistical. Together with the simpler proof reported in Ref.~\cite{Davis2022b} where it was shown that superstatistics \emph{propagates downwards}, 
namely that if $AB$ is superstatistical then so is $A$, this work establishes a closure property of superstatistics.

\section{Steady states and superstatistics}

Consider a system with microstates $\bm \Gamma \in \mathcal{V}$ and Hamiltonian $\H(\bm \Gamma)$ in a steady state where the microstate distribution is of the form
\begin{equation}
\label{eq:rho}
P(\bm \Gamma|S) = \rho\big(\H(\bm \Gamma); S\big),
\end{equation}
with $\rho$ the \emph{ensemble function}. A particular case of steady states are given by superstatistics~\cite{Beck2003, Beck2004}, a framework where the inverse temperature $\beta \defeq 1/(k_B T)$ 
is a random variable, with probability density $P(\beta|S)$ and joint density
\begin{equation}
P(\bm \Gamma, \beta|S) = P(\beta|S)\frac{\exp\big(-\beta \H(\bm \Gamma)\big)}{Z(\beta)}.
\end{equation}

\noindent
By integrating out $\beta$ we obtain the superstatistical microstate distribution
\begin{equation}
P(\bm \Gamma|S) = \int_0^\infty d\beta\,P(\beta|S)\frac{\exp\big(-\beta \H(\bm \Gamma)\big)}{Z(\beta)},
\end{equation}
which has the form in \eqref{eq:rho} with ensemble function
\begin{equation}
\label{eq:laplace}
\rho(E; S) = \int_0^\infty d\beta\,f(\beta; S)\exp(-\beta E),
\end{equation}
where we see that $\rho$ is the Laplace transform of a weight function
\begin{equation}
\label{eq:weight}
f(\beta; S) \defeq \frac{P(\beta|S)}{Z(\beta)}.
\end{equation}

Notice that the form of the ensemble function, and therefore of the microstate distribution, is completely determined by the weight function $f(\beta; S)$ in 
\eqref{eq:weight} through the Laplace transform in \eqref{eq:laplace}. Therefore, $\rho$ depends on both the inverse temperature distribution and the partition function of the system, and this implies 
that two systems with the same $P(\beta|S)$ can be described by completely different ensembles, depending on their respective partition functions, or equivalently, densities of states. The sole exception 
is the case of the canonical ensemble, where $P(\beta|\beta_0) = \delta(\beta-\beta_0)$ and the resulting ensemble function is always $\rho(E; \beta_0) \propto \exp(-\beta_0 E)$ regardless of the form of 
the partition function.

\noindent
The fundamental inverse temperature function~\cite{Davis2019, Davis2023b} of the ensemble, defined by
\begin{equation}
\label{eq:fundtemp}
\beta_F(E; S) \defeq -\frac{\partial}{\partial E}\ln \rho(E; S),
\end{equation}
is such that
\begin{equation}
\beta_F(E; S) = \big<\beta\big>_{E, S} = \int_0^\infty d\beta\,P(\beta|E, S)\beta,
\end{equation}
with $P(\beta|E, S)$ the conditional inverse temperature distribution, given by
\begin{equation}
P(\beta|E, S) = \frac{P(E|\beta)P(\beta|S)}{P(E|S)} = \frac{\exp(-\beta E)f(\beta; S)}{\rho(E; S)}.
\end{equation}

\section{Closure property of superstatistics}

Let us consider a composite system with degrees of freedom $\G \defeq (\Ga; \Gb)$, that is, divided into two subsystems, one with degrees of freedom $\Ga \in \mathcal{V}_A$ and 
another with degrees of freedom $\Gb \in \mathcal{V}_B$. Furthermore, we will assume that the Hamiltonian $\H(\G)$ is of the form
\begin{equation}
\label{eq:additive}
\H(\G) = \H(\Ga, \Gb) = \H_A(\Ga) + \H_B(\Gb),
\end{equation}
where $\H_A$ and $\H_B$ are the Hamiltonians for the subsystems $A$ and $B$, respectively. 

In a steady state $S$ described by an ensemble function $\rho(E; S)$, the joint distribution of microstates for the composite system is given by
\begin{equation}
P(\Ga, \Gb|S) = \rho\big(\H_A(\Ga)+\H_B(\Gb); S\big),
\end{equation}
and then the marginal distribution for the subsystem $A$ is
\begin{equation}
\label{eq:marginal_pre}
P(\Ga|S) = \IGb\,P(\Ga, \Gb|S) = \IGb\,\rho\big(\H_A(\Ga) + \H_B(\Gb); S).
\end{equation}

Recognizing that the rightmost integral in \eqref{eq:marginal_pre} depends on $\Ga$ only through $\H_A(\Ga)$, we see that $\Ga$ is also in a steady state with microstate distribution
\begin{equation}
P(\Ga|S) = \rho_A\big(\H_A(\Ga); S\big),
\end{equation}
and where the ensemble function $\rho_A$ is given by
\begin{equation}
\rho_A(\varepsilon; S) \defeq \IGb\,\rho(\varepsilon+\H_B(\Gb); S).
\end{equation}

In general, $\rho_A$ belongs to a different family than $\rho$, a notable exception being, as is well known, the canonical ensemble, where
\begin{equation}
P(\Ga, \Gb|\beta) = \frac{\exp\Big(-\beta\big[\H_A(\Ga)+\H_B(\Gb)\big]\Big)}{Z_{AB}(\beta)}.
\end{equation}

\noindent
In that case we have
\begin{equation}
P(\Ga, \Gb|\beta) = P(\Ga|\beta) P(\Gb|\beta)
\end{equation}
with
\begin{align}
P(\bm{\Gamma}_\alpha|\beta) & = \frac{\exp\big(-\beta \H_{\alpha}(\bm{\Gamma}_\alpha)\big)}{Z_{\alpha}(\beta)}, \\
Z_{\alpha}(\beta) & \defeq \int_{\mathcal{V}_{\alpha}} d\bm{\Gamma}_{\alpha}\,\exp\big(-\beta \H_{\alpha}(\bm{\Gamma}_\alpha)\big)
\end{align}
for $\alpha = A, B$ and
\begin{equation}
Z_{AB}(\beta) = Z_A(\beta) Z_B(\beta).
\end{equation}

In other words, canonical composite systems have canonical subsystems, and is natural to look for a generalization of this statement to superstatistics. As it turns out, the 
following lemma is true. 

\begin{lemma}[Downwards propagation of superstatistics]
\label{lem:1}

\phantom{.}\\[3pt] If the joint steady-state ensemble given by
\begin{equation}
P(\Ga, \Gb|S) = \rho\big(\H_A(\Ga) + \H_B(\Gb); S\big)
\end{equation}
is a superstatistical ensemble with inverse temperature distribution $P(\beta|S)$, then the marginal ensemble
\begin{equation}
P(\Ga|S) = \IGb P(\Ga, \Gb|S)
\end{equation}
is also a superstatistical ensemble with the same $P(\beta|S)$.
\end{lemma}

The proof is simple~\cite{Davis2022b, Davis2024b}, so is reproduced here for convenience. Assume that our composite system $\G$ is superstatistical, that is,
\begin{equation}
\label{eq:super_AB}
P(\Ga, \Gb|S) = \int_0^\infty d\beta P(\beta|S)P(\Ga, \Gb|\beta).
\end{equation}

\noindent
The distribution for $\Ga$ is obtained by integrating over $\Gb$, yielding
\begin{equation}
\begin{split}
P(\Ga|S) & = \IGb\int_0^\infty d\beta P(\beta|S)P(\Ga, \Gb|\beta) \\
& = \IGb\int_0^\infty d\beta P(\beta|S)P(\Ga|\beta)P(\Gb|\beta) \\
& = \int_0^\infty d\beta P(\beta|S)P(\Ga|\beta)\IGb\,P(\Gb|\beta),
\end{split}
\end{equation}
but the rightmost integral in the last line is equal to 1 because $P(\Gb|\beta)$ is normalized, therefore we have
\begin{equation}
\label{eq:super_A}
P(\Ga|S) = \int_0^\infty d\beta P(\beta|S)P(\Ga|\beta),
\end{equation}
which means $\Ga$ is also described by the superstatistical ensemble generated by $P(\beta|S)$.

On the other hand, the fact that \eqref{eq:super_A} also implies \eqref{eq:super_AB} is a much stringent property whose proof, presented in \ref{sec:appendix1}, constitutes the 
main result of the present work. It is stated in the following theorem.

\begin{theorem}[Upwards propagation of superstatistics]
\label{thm:the_theorem}

The unique solution for $P(\Ga, \Gb|S)$ of the marginalization equation 
\begin{equation}
\label{eq:functional}
\IGb\,P(\Ga, \Gb|S) = \int_0^\infty d\beta\,P(\beta|S)P(\Ga|\beta)
\end{equation}
of the form
\begin{equation}
\label{eq:rho_AB}
P(\Ga, \Gb|S) = \rho\big(\H_A(\Ga)+\H_B(\Gb); S\big)
\end{equation}
is
\begin{equation}
P(\Ga, \Gb|S) = \int_0^\infty d\beta\,P(\beta|S)P(\Ga, \Gb|\beta),
\end{equation}
regardless of the choice of $\mathcal{V}_B$ and $\H_B$.

\end{theorem}

Taken together, Lemma~\ref{lem:1} and Theorem~\ref{thm:the_theorem} constitute a closure property of superstatistics for additive Hamiltonians, that is, where \eqref{eq:additive} is valid, in the sense 
that both removing and adding degrees of freedom via marginalization preserves the superstatistical nature of the steady-state ensemble. It is important to note that the inverse temperature distribution 
$P(\beta|S)$ is also preserved when adding or removing degrees of freedom.

Because the fundamental inverse temperature, defined in \eqref{eq:fundtemp}, for a marginal ensemble is given by~\cite{Davis2022b}
\begin{equation}
\beta_F^{(A)}(E_A; S) = \Big<\beta_F(E; S)\Big>_{E_A, S} \quad,
\end{equation}
with $E = E_A + E_B$, an alternative formulation of Theorem \ref{thm:the_theorem} is that the only solution for $\beta_F(E; S)$ of the equation
\begin{equation}
\label{eq:eq}
\big<\beta_F\big>_{E_A, S} = \big<\beta\big>_{E_A, S}
\end{equation}
is actually
\begin{equation}
\label{eq:sol}
\beta_F(E; S) = \big<\beta\big>_{E, S}.
\end{equation}

In order to verify that \eqref{eq:sol} is solution of \eqref{eq:eq}, we simply replace it in the left-hand side of \eqref{eq:eq} and rearrange the integrals, obtaining
\begin{equation}
\begin{split}
\Big<\big<\beta\big>_{E, S}\Big>_{E_A, S} & = \int_0^\infty dE\,P(E|E_A, S)\big<\beta\big>_{E, S} \\
& = \int_0^\infty dE\,P(E|E_A, S)\left[\int_0^\infty d\beta\,P(\beta|E, S)\beta\right] \\
& = \int_0^\infty d\beta\,\left[\int_0^\infty dE\,P(\beta|E, S)P(E|E_A, S)\right]\beta \\
& = \int_0^\infty d\beta\,P(\beta|E_A, S)\beta = \big<\beta\big>_{E_A, S}.
\end{split}
\end{equation}

This is the equivalent of Lemma~\ref{lem:1}, but written in terms of expectation values of inverse temperatures. In the following sections we will discuss several consequences of 
Theorem~\ref{thm:the_theorem} and Lemma~\ref{lem:1}.

\section{No local thermal equilibrium in non-canonical steady states}

Let us now state the idea of local thermal equilibrium (LTE) in a non-canonical steady state. LTE is the existence of canonical ensembles describing a subset of the degrees of freedom of the system, while 
the composite system is not canonical. In some contexts, the superposition of canonical ensembles that is postulated by superstatistics itself is understood as a collection of regions in LTE at different 
temperatures.

It is simple to prove, using the closure property, that for a system of classical particles in a non-canonical steady state where the only interactions are with an external potential, LTEs cannot exist.
For the proof, we will consider a system of $N$ classical particles with microstates
\begin{equation}
\bm \Gamma \defeq (\bm{r}_1, \ldots, \bm{r}_N, \bm{p}_1, \ldots, \bm{p}_N),
\end{equation}
which, without loss of generality, we will divide into two groups or subsystems: the first where we will postulate LTE, composed of $L$ particles and a second with $N-L$ particles. Because the only 
interactions present are with the external potential, the Hamiltonian of the full system can be written as
\begin{equation}
\H(\bm \Gamma) = \sum_{i=1}^L \left[\frac{\bm{p}_i^2}{2 m_i} + \phi_i\big(\bm{r}_i\big) \right] + \sum_{i=L+1}^N \left[\frac{\bm{p}_i^2}{2 m_i} + \phi_i\big(\bm{r}_i\big) \right].
\end{equation}

\noindent
Now, let us assume that the subsystem $A$ with $L$ particles and 
\begin{equation}
\Ga \defeq (\bm{r}_1, \ldots, \bm{r}_L, \bm{p}_1, \ldots, \bm{p}_L)
\end{equation}
is in LTE at inverse temperature $\beta_0$. We would have
\begin{equation}
\label{eq:localeq}
P(\bm{p}_1, \ldots, \bm{p}_L, \bm{r}_1, \ldots, \bm{r}_L|\beta_0) = \frac{1}{Z_L(\beta_0)}\exp\left(-\beta_0\sum_{i=1}^L \left[\frac{\bm{p}_i^2}{2 m_i} + \phi_i\big(\bm{r}_i\big) \right]\right)
\end{equation}
hence the subsystem is canonical, and therefore superstatistical with \[P(\beta|\beta_0) = \delta(\beta-\beta_0).\]

The full, $N$-particle system must also be canonical, according to Theorem \ref{thm:the_theorem}, and we must have
\begin{equation}
P(\bm \Gamma|S) = \int_0^\infty d\beta\,\delta(\beta-\beta_0)\,P(\bm \Gamma|\beta) = \frac{1}{Z(\beta_0)}\exp\left(-\beta_0\sum_{i=1}^N \left[\frac{\bm{p}_i^2}{2 m_i} + \phi_i\big(\bm{r}_i\big) \right]\right)
\end{equation}
which contradicts the assumption of a non-canonical steady state. We have then that the condition of local equilibrium in \eqref{eq:localeq} must be false.

\section{Non-canonical steady states cannot have Maxwellian velocity distributions}

Because LTE in terms of velocities corresponds to Maxwellian velocity distributions, we have a direct parallel of the proof of impossibility of LTEs: a system in a non-canonical steady state cannot 
have Maxwellian velocity distributions for its particles. This can be understood as a generalization of the result by Ray and Graben~\cite{Ray1991b}, where systems of classical particles in the microcanonical 
ensemble are found to have non-Maxwellian distributions in general, and the Maxwellian distribution is only recovered in the thermodynamic limit (i.e. infinite degrees of freedom). The proof of this statement 
is rather straightforward, being a parallel of the last proof except we can now allow for arbitrary interactions between the particles. First, we take
\begin{equation}
\bm \Gamma \defeq ( \bm{p}_1, \ldots, \bm{p}_N, \bm{r}_1, \ldots, \bm{r}_N)
\end{equation}
and write the Hamiltonian as
\begin{equation}
\H(\bm \Gamma) = \frac{\bm{p}_1^2}{2 m_1} + \left[\sum_{i=2}^N \frac{\bm{p}_i^2}{2 m_i} + \Phi\big(\bm{r}_1, \ldots, \bm{r}_N\big)\right],
\end{equation}
such that we divide $\bm \Gamma$ into two subsystems, subsystem $A$ having $\Ga \defeq \bm{p}_1$ and subsystem $B$ with $\Gb \defeq (\bm{p}_2, \ldots, \bm{p}_N, \bm{r}_1, \ldots, \bm{r}_N)$.
Then we assume that a non-canonical steady state $P(\bm \Gamma|S)$ exists such that the marginal distribution for $\Ga$, i.e. $\bm{p}_1$, follows a Maxwellian distribution. Thus we have
\begin{equation}
\label{eq:coro_maxwell}
P(\bm{p}_1|S) = \int d\Gb P(\bm{p}_1, \Gb|S) = \left(\frac{\beta_0}{2\pi m_1}\right)^{\frac{3}{2}}\exp\left(-\frac{\beta_0\bm{p}_1^2}{2 m_1}\right).
\end{equation}

But this is a canonical ensemble, therefore superstatistical with $P(\beta|\beta_0) = \delta(\beta-\beta_0)$, and by Theorem \ref{thm:the_theorem}, we must have
\begin{equation}
\label{eq:coro_canon}
\begin{split}
P(\bm{p}_1, \Gb|S) & = \int_0^\infty d\beta \delta(\beta-\beta_0) P(\bm{p}_1, \Gb|\beta) \\
& = \frac{1}{Z(\beta_0)}\exp\left(-\beta_0\left[\sum_{i=1}^N \frac{\bm{p}_i^2}{2 m_i} + \Phi(\bm{r}_1, \ldots, \bm{r}_N)\right]\right).
\end{split}
\end{equation}
 
As this is also a canonical ensemble, it contradicts the assumption that $S$ was a non-canonical steady state, and it follows that \eqref{eq:coro_maxwell} is false.

\newpage
\section{Spatial distribution of particles in collisionless plasmas}

Following Ref.~\cite{Davis2019b}, for a collisionless plasma under a self-consistent electrostatic potential $V(\bm r)$ and a vector potential $\bm{A}(\bm r)$, such that the Hamiltonian is given by
\begin{equation}
\label{eq:plasma_ham}
\H(\bm{r}_1, \ldots, \bm{r}_N, \bm{p}_1, \ldots, \bm{p}_N) = \sum_{i=1}^N \left[\frac{\big(\bm{p}_i-q_i\bm{A}(\bm{r}_i)\big)^2}{2 m_i} + q_i V(\bm{r}_i)\right]
\end{equation}
the probability of microstates $(\bm{r}_1, \ldots, \bm{r}_N, \bm{v}_1, \ldots, \bm{v}_N)$ can be written as
\begin{equation}
P(\bm{r}_1, \ldots, \bm{r}_N, \bm{v}_1, \ldots, \bm{v}_N|S) = \rho\Big(\mathcal{E}(\bm{r}_1, \ldots, \bm{r}_N, \bm{v}_1, \ldots, \bm{v}_N); S\Big)
\end{equation}
where
\begin{equation}
\mathcal{E}(\bm{r}_1, \ldots, \bm{r}_N, \bm{v}_1, \ldots, \bm{v}_N) = \sum_{i=1}^N \left[\frac{m_i \bm{v}_i^2}{2} + q_i V\big(\bm{r}_i\big)\right]
\end{equation}
is the energy function, that is, the Hamiltonian in \eqref{eq:plasma_ham} but written in terms of particle velocities instead of particle momenta. In thermal equilibrium the particle velocities follow 
the Maxwellian distribution,
\begin{equation}
P(\bm{v}_i = \bm{v}|\beta) = \left(\frac{m_i \beta}{2\pi}\right)^{\frac{3}{2}}\exp\left(-\frac{\beta m_i \bm{v}^2}{2}\right),
\end{equation}
and, according to the result in Ref.~\cite{Davis2019b}, the only admissible single-particle distributions in collisionless plasmas are described by superstatistics. Therefore we can 
write, for a steady state $S$,
\begin{equation}
P(\bm{v}_i = \bm{v}|S) = \left(\frac{m_i}{2\pi}\right)^{\frac{3}{2}}\int_0^\infty d\beta\,P(\beta|S)\,\beta^{\frac{3}{2}}\exp\left(-\frac{\beta m_i\bm{v}^2}{2}\right),
\end{equation}
for some distribution $P(\beta|S)$. Theorem \ref{thm:the_theorem} then implies, for the entire microstate, that
\begin{equation}
P(\bm{r}_1, \ldots, \bm{r}_N, \bm{v}_1, \ldots, \bm{v}_N|S) = \int_0^\infty d\beta\,\frac{P(\beta|S)}{Z(\beta)}\exp\left(-\beta \sum_{i=1}^N\left[\frac{m_i\bm{v}_i^2}{2} + q_i V(\bm{r}_i)\right]\right),
\end{equation}
and it follows by integration over all $\bm{v}$ and all $\bm{r}$ except $\bm{r}_i$ that
\begin{equation}
\label{eq:spatial}
P(\bm{r}_i|S) = \int_0^\infty d\beta\,P(\beta|S)\frac{\exp\big(-\beta q_i V(\bm{r}_i)\big)}{Z_1(\beta; q_i)}.
\end{equation}
with
\begin{equation}
Z_1(\beta; q) \defeq \int d\bm{r}\,\exp\big(-\beta q V(\bm r)\big).
\end{equation}

The electrostatic potential $V$ is then closely related to the spatial distribution of the particles. As an example, consider particles with kappa distributions for their velocities and where the 
particles are enclosed in a box such that \[\bm{r} \in [-L/2, L/2]\times [-L/2, L/2] \times [-L/2, L/2]\] and the electrostatic potential inside is
\begin{equation}
V(\bm r) = - E_0 x,
\end{equation}
so that they are subject to a constant electric field $\bm{E} = -\nabla V = E_0 \hat{\bm X}$ in the $X$ direction. The inverse temperature distribution given by a gamma distribution,
\begin{equation}
\label{eq:gamma}
P(\beta|u, \beta_S) = \frac{1}{u\beta_S\,\Gamma(1/u)}\exp\left(-\frac{\beta}{u\beta_S}\right)\left(\frac{\beta}{u\beta_S}\right)^{\frac{1}{u}-1},
\end{equation}
leads to the kappa distribution~\cite{Davis2023e} in the form,
\begin{equation}
P(\bm{v}|u, \beta_S) = \left(\frac{m u\beta_S}{2\pi}\right)^{\frac{3}{2}}\frac{\Gamma\left(\frac{3}{2}+\frac{1}{u}\right)}{\Gamma\left(\frac{1}{u}\right)}
\Big[1+ u\beta_S\frac{m\bm{v}^2}{2}\Big]^{-\left(\frac{1}{u}+\frac{3}{2}\right)}
\end{equation}
where
\begin{equation}
\label{eq:ukappa}
\kappa \defeq \frac{1}{u}+\frac{1}{2}.
\end{equation}

\noindent
The partition function $Z_1(\beta; q)$ is given in this case by
\begin{equation}
\label{eq:part}
Z_1(\beta; q) = \int_{-\frac{L}{2}}^{\frac{L}{2}} dx \int_{-\frac{L}{2}}^{\frac{L}{2}} dy \int_{-\frac{L}{2}}^{\frac{L}{2}} dz \exp(\beta q E_0 x)
= \frac{2 L^2}{\big|q\big|\beta E_0}\text{sinh}\left(\frac{\big|q\big|\beta E_0 L}{2}\right),
\end{equation}
so replacing \eqref{eq:gamma} and \eqref{eq:part} into \eqref{eq:spatial} we have
\begin{equation}
\begin{split}
P(\bm{r}_i|u, \beta_S) & = \int_0^\infty d\beta\,P(\beta|S)\left[\frac{|q_i|\beta E_0}{2L^2}\frac{\exp\left(\beta q_i E_0 x_i\right)}{\text{sinh}\Big(\frac{|q_i|\beta E_0L}{2}\Big)}\right] \\
& = \frac{\big|q_i\big|E_0}{2L^2\,\Gamma(1/u)}\int_0^\infty d\beta\,\left[\frac{\exp\left(-\frac{\beta}{u\beta_S} + \beta q_i E_0 x_i\right)}{\text{sinh}\Big(\frac{\big|q_i\big|\beta E_0 L}{2}\Big)}\right]
\left(\frac{\beta}{u\beta_S}\right)^{\frac{1}{u}}.
\end{split}
\end{equation}

\noindent
By defining the dimensionless parameter 
\begin{equation}
R \defeq u\beta_S E_0 L \big|q_i\big|
\end{equation}
we can write
\begin{equation}
P(\bm{r}_i|u, \beta_S) = \frac{R}{2L^3\,\Gamma(1/u)}\int_0^\infty dt\,\exp\left(-t\Big[1 - R\,\text{sgn}(q_i)\Big(\frac{x_i}{L}\Big)\Big]\right) t^{\frac{1}{u}}
\;\text{sinh}\left(\frac{R t}{2}\right)^{-1}
\end{equation}
and, using the integral
\begin{equation}
\int_0^\infty d\beta\,\left[\frac{\exp(-\beta/\theta)\beta^{k-1}}{\text{sinh}(\beta c)}\right] = \frac{2\,\Gamma(k)}{(2c)^k} \zeta\Big(k, \frac{1}{2}\Big[1+\frac{1}{c\theta}\Big]\Big)
\end{equation}
for $k > 1$ and $0 < c < 1/\theta$ where $\zeta(z, a)$ is the Hurwitz zeta function, we finally obtain the spatial distribution as a function of $u$ and $\beta_S$,
\begin{equation}
\label{eq:posdist}
P(\bm{r}_i|u, \beta_S) = \frac{1}{u L^3\,R^{\frac{1}{u}}}\,\zeta\left(\frac{1}{u}+1, \frac{1}{R} -\text{sgn}(q_i)\frac{x_i}{L} + \frac{1}{2}\right),
\end{equation}
such that, in the limit $u \rightarrow 0$ and $\beta_S \rightarrow \beta$ it reduces to the canonical distribution
\begin{equation}
P(\bm{r}_i|\beta) = \frac{|q_i|\beta E_0}{2L^2}\frac{\exp\left(\beta q_i E_0 x_i\right)}{\text{sinh}\Big(\frac{|q_i|\beta E_0 L}{2}\Big)}.
\end{equation}

\begin{figure}
\begin{center}
\includegraphics[width=0.49\textwidth]{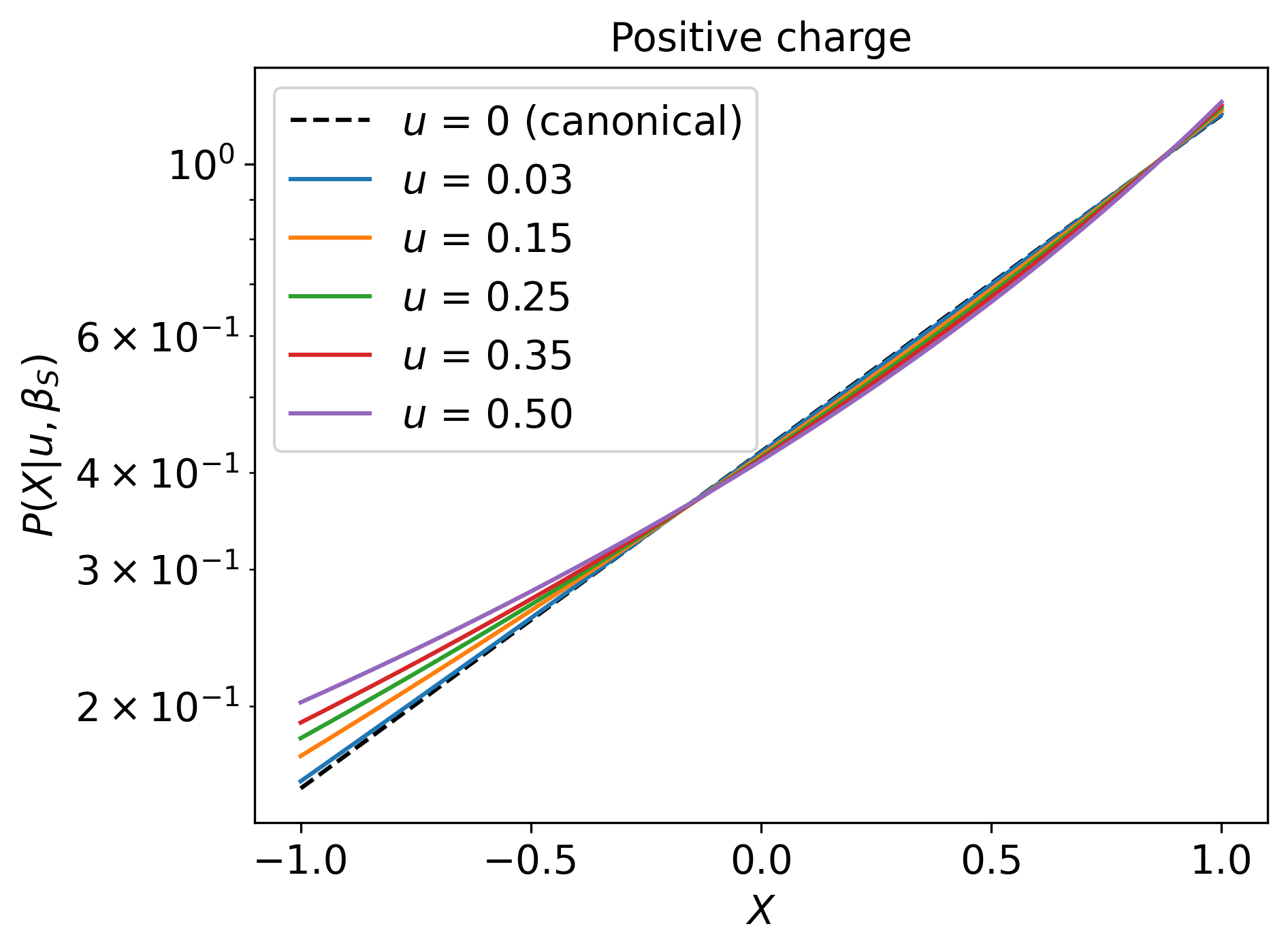}
\includegraphics[width=0.49\textwidth]{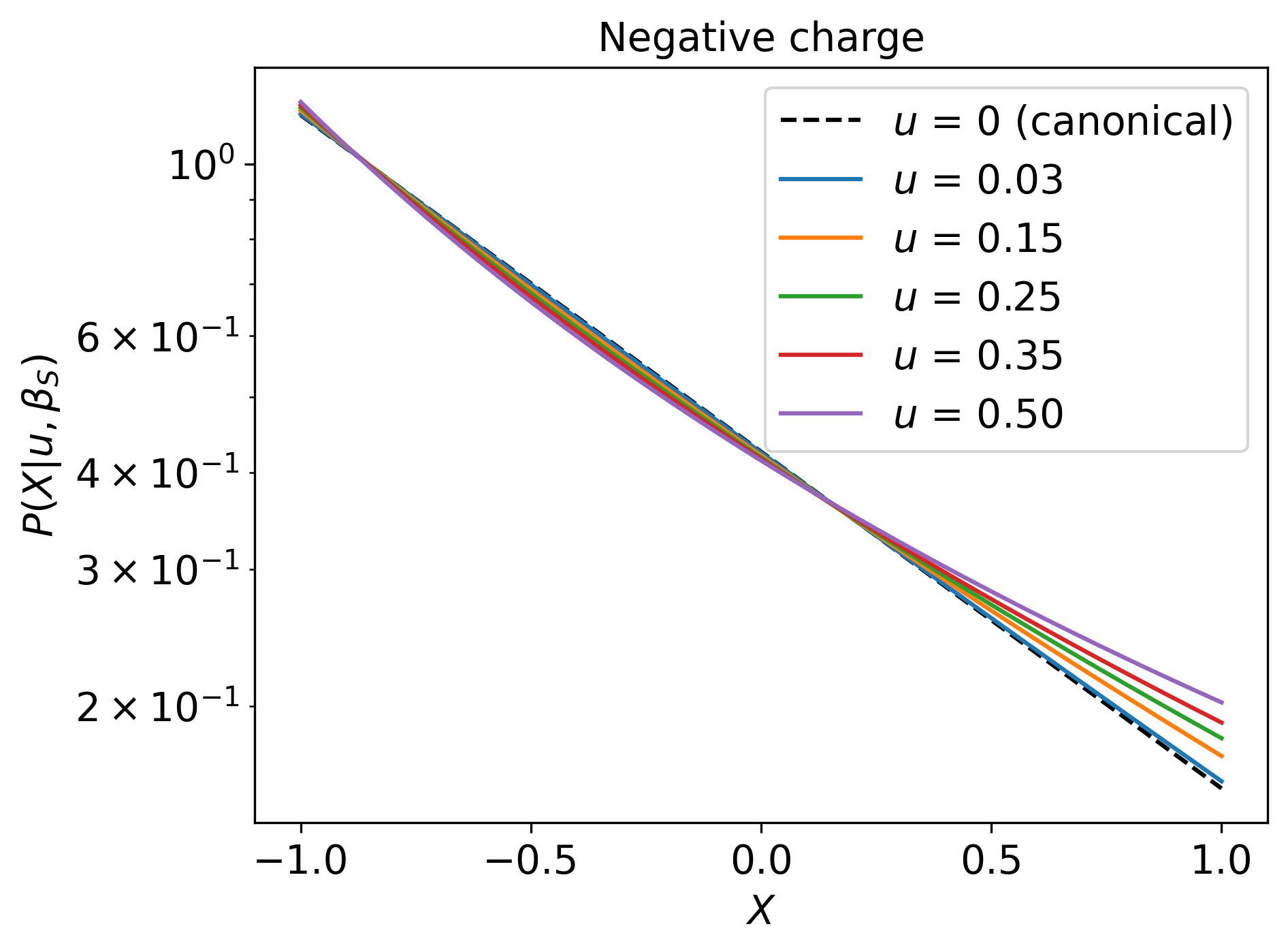}
\end{center}
\caption{Position distribution in \eqref{eq:posdist} for positive and negative charged particles, obtained using $P(\beta|u, \beta_S)$ given by a gamma distribution. Increasing values of $u$ correspond 
to decreasing values of the spectral index $\kappa$, according to \eqref{eq:ukappa}.}
\label{fig:posdist}
\end{figure}

\newpage
Fig.~\ref{fig:posdist} shows the position distribution in \eqref{eq:posdist} for positive and negative charges. As expected, we see that the particles concentrate at the right wall ($X = L/2$) for 
positive charges, that is, they move in the direction of the electric field, while for negative charges they move against the field, concentrating at the left wall ($X = -L/2$). We also see that the 
tails of the spatial distribution are longer for increasing $u$, a parallel of what is observed in the tails of the kappa distribution.

Notice that the distribution in \eqref{eq:posdist} does not belong to the $q$-exponential family, as is the case for the kappa distribution of velocities. We in fact obtain completely different 
distributions from the same $P(\beta|S)$ due to the radically different forms of the partition functions of the two subsystems. This is in contradiction with a common assumption where kappa distributions 
of velocities are always postulated together with kappa-like power laws (in effect, $q$-exponential distributions) for potential energy or configurational degrees of freedom~\cite{Livadiotis2015, 
Livadiotis2016, Livadiotis2017b}, regardless of the functional form of the partition functions or densities of state of the subsystems involved.

\section{Concluding remarks}

We have shown that superstatistics propagates upwards, that is, the existence of superstatistics for a subsystem $A$ implies superstatistics for the composite system $AB$. Together with the fact, proved 
earlier, that superstatistics also propagates downwards, it follows that a system is superstatistical if and only if each of its additive parts are also superstatistical. We have presented some implications 
of this theorem for classical particle systems, regarding the impossibility of local thermal equilibrium and Maxwellian velocity distributions for regions of system in a non-canonical steady state.

\newpage
Despite the fact that the superstatistical inverse temperature distributions for additive subsystems are the same, and are also equal to that of the composite system, it is possible to have different 
microstate distributions, and therefore, different fundamental inverse temperatures for $A$, $B$ and $AB$, depending on the corresponding partition functions.

\section*{Acknowledgments}

Funding from ANID FONDECYT 1220651 grant is gratefully acknowledged.

\appendix
\section{Proof of the main theorem}
\label{sec:appendix1}

We start from the functional equation \eqref{eq:functional}, replacing $P(\Ga|\beta)$ according to the canonical ensemble,
\begin{equation}
P(\Ga|\beta) = \frac{\exp\big(-\beta \H_A(\Ga)\big)}{Z_A(\beta)},
\end{equation}
and also replacing $P(\Ga, \Gb|S)$ using \eqref{eq:rho_AB}. The equation now has the form
\begin{equation}
\label{eq:functional_appendix}
\int_0^\infty d\beta\,P(\beta|S)\frac{\exp\big(-\beta \H_A(\Ga)\big)}{Z_A(\beta)} = \IGb\,\rho\big(\H_A(\Ga) + \H_B(\Gb); S).
\end{equation}

\noindent
Letting $\varepsilon \defeq \H_A(\Ga)$ and introducing the density of states of the environment
\begin{equation}
\Omega_B(E_B) \defeq \IGb\,\delta\big(\H_B(\Gb)-E_B\big)
\end{equation}
in the right-hand side, we can write \eqref{eq:functional_appendix} as
\begin{equation}
\begin{split}
\int_0^\infty d\beta\,f_A(\beta; S)\exp(-\beta \varepsilon) & = \int_0^{\infty} dE_B\,\Omega_B(E_B)\,\rho(\varepsilon + E_B; S) \\
& = \int_{-\infty}^{\infty} dE\,\Omega_B(E-\varepsilon)\Theta(E-\varepsilon)\,\rho(E; S),
\end{split}
\end{equation}
where we have applied the change of variables $E \defeq \varepsilon + E_B$ to the right-hand side, and we have defined
\begin{equation}
f_A(\beta; S) \defeq \frac{P(\beta|S)}{Z_A(\beta)}.
\end{equation}

Now we take the Fourier transform on both sides with respect to the variable $\varepsilon$, that is, for a function $g(\varepsilon)$ we construct $\tilde{g}(u)$ according to
\begin{equation}
\tilde{g}(u) \defeq \frac{1}{\sqrt{2\pi}}\int_{-\infty}^{\infty} d\varepsilon\,g(\varepsilon)\exp(\iu u \varepsilon),
\end{equation}
that is, we have
\begin{equation}
\label{eq:foo}
\begin{split}
\int_{-\infty}^{\infty} d\varepsilon \frac{\exp(\iu u \varepsilon)}{\sqrt{2\pi}}& \left[\int_0^\infty d\beta\,f_A(\beta; S)\exp(-\beta \varepsilon)\right] \\
& = \int_{-\infty}^{\infty} d\varepsilon\,\frac{\exp(\iu u \varepsilon)}{\sqrt{2\pi}}\left[\int_{-\infty}^{\infty} dE\,\Omega_B(E-\varepsilon)\Theta(E-\varepsilon)\,\rho(E; S)\right].
\end{split}
\end{equation}

\noindent
Because the left-hand side of \eqref{eq:foo} can be rearranged as
\begin{equation}
\int_0^\infty d\beta\,f_A(\beta; S)\left[\int_{-\infty}^{\infty} d\varepsilon \frac{\exp\big([\iu u -\beta]\varepsilon\big)}{\sqrt{2\pi}}\right] = \sqrt{2\pi}f_A(\iu u; S),
\end{equation}
we can write \eqref{eq:foo} in the form
\begin{equation}
\begin{split}
f_A(\iu u; S) & = \int_{-\infty}^{\infty} d\varepsilon \int_{-\infty}^{\infty} dE\,\Omega_B(E-\varepsilon)\Theta(E-\varepsilon)\,\rho(E; S)\frac{\exp(\iu u \varepsilon)}{2\pi} \\
& = \left[\int_{-\infty}^{\infty} dz\,\Omega_B(-z)\Theta(-z)\frac{\exp(\iu uz)}{\sqrt{2\pi}}\right]\cdot \left[\int_{-\infty}^{\infty} dE\,\rho(E; S)\frac{\exp(\iu u E)}{\sqrt{2\pi}}\right],
\end{split}
\end{equation}
that is,
\begin{equation}
\label{eq:f1}
f_A(\iu u; S) = \tilde{W}_B(u)\tilde{\rho}(u; S),
\end{equation}
where we have defined
\begin{align}
\label{eq:rho_tilde}
\tilde{\rho}(u; S) & \defeq \frac{1}{\sqrt{2\pi}}\int_{-\infty}^{\infty} dE\,\rho(E; S)\exp(\iu u E), \\
\label{eq:W2_tilde}
\tilde{W}_B(u) & \defeq \frac{1}{\sqrt{2\pi}}\int_{-\infty}^{\infty} dz\,\Omega_B(-z)\Theta(-z)\exp(\iu uz).
\end{align}

\noindent
Taking the inverse Fourier transform of \eqref{eq:rho_tilde} and using \eqref{eq:f1} we have
\begin{equation}
\label{eq:rho2}
\rho(E; S) = \frac{1}{\sqrt{2\pi}} \int_{-\infty}^{\infty} du \exp(-\iu u E)\frac{f_A(\iu u; S)}{\tilde{W}_B(u)}
\end{equation}
and writing
\begin{equation}
f_A(\iu u; S) = \int_0^\infty d\beta\,f_A(\beta; S)\delta(\beta-\iu u)
\end{equation}
we can rewrite \eqref{eq:rho2} as
\begin{equation}
\begin{split}
\rho(E; S) & = \int_0^\infty d\beta\,f_A(\beta; S)\left[\int_{-\infty}^{\infty} du\,\frac{\exp(-\iu u E)}{\sqrt{2\pi}\,\tilde{W}_B(u)}\delta\big(\beta-\iu u\big)\right] \\
& = \int_0^\infty d\beta\,f_A(\beta; S)\frac{\exp(-\beta E)}{\sqrt{2\pi}\tilde{W}_B(-\iu \beta)}
\end{split}
\end{equation}
but from \eqref{eq:W2_tilde} we have
\begin{equation}
\begin{split}
\sqrt{2\pi}\tilde{W}_B(-\iu \beta) & = \int_{-\infty}^{\infty} dz\,\Omega_B(-z)\Theta(-z)\exp(\beta z) \\
& = \int_0^{\infty} dE_B\,\Omega(E_B)\exp(-\beta E_B) = Z_B(\beta)
\end{split}
\end{equation}
thus we have
\begin{equation}
\rho(E; S) = \int_0^\infty d\beta\,P(\beta|S)\frac{\exp(-\beta E)}{Z_A(\beta)Z_B(\beta)},
\end{equation}
or equivalently,
\begin{equation}
P(\Ga, \Gb|S) = \int_0^\infty d\beta\,P(\beta|S)P(\Ga, \Gb|\beta).
\end{equation}

\section*{References}

\bibliography{closure}
\bibliographystyle{unsrt}
\end{document}